\documentclass[%
 reprint,
 twocolumn,amsmath, amssymb, aps,superscriptaddress
]{revtex4-2}

\usepackage{graphicx}
\usepackage{bm}
\usepackage{physics}
\usepackage{xcolor}
\usepackage[normalem]{ulem}
\usepackage{hyperref}
\usepackage{lipsum}  
\usepackage{tikz}
\usepackage{comment}
 \usepackage{cleveref}
\newif\ifstartedinmathmode
\newcommand\encircled[1]{%
  \relax\ifmmode\startedinmathmodetrue\else\startedinmathmodefalse\fi%
  \tikz[baseline,anchor=base]{%
  \node[draw,circle,outer sep=0pt,inner sep=.2ex]
    {\ifstartedinmathmode$#1$\else#1\fi};}%
}

\definecolor{darkblue}{rgb}{0.0,0.0,0.4}
\definecolor{darkgreen}{rgb}{0.0,0.25,0.0}
\definecolor{shadecolor}{gray}{1}
\definecolor{markus}{rgb}{0.09,0.65,0.63}
\definecolor{marius}{rgb}{1,0.5,0.0}
\definecolor{elinor}{rgb}{0.3,0.4,1.0}

\makeindex

\begin{document}

\title{Observation of supersolid-like sound modes in a driven quantum gas}

\author{Nikolas Liebster}
\email{superfluid-crystal@matterwave.de}
\affiliation{Kirchhoff-Institut f\"{u}r Physik, Universit\"{a}t Heidelberg, Im Neuenheimer Feld 227, 69120 Heidelberg, Germany}
\author{Marius Sparn}
\affiliation{Kirchhoff-Institut f\"{u}r Physik, Universit\"{a}t Heidelberg, Im Neuenheimer Feld 227, 69120 Heidelberg, Germany}
\author{Elinor Kath}
\affiliation{Kirchhoff-Institut f\"{u}r Physik, Universit\"{a}t Heidelberg, Im Neuenheimer Feld 227, 69120 Heidelberg, Germany}
\author{Jelte Duchene}
\affiliation{Kirchhoff-Institut f\"{u}r Physik, Universit\"{a}t Heidelberg, Im Neuenheimer Feld 227, 69120 Heidelberg, Germany}
\author{Helmut Strobel}
\affiliation{Kirchhoff-Institut f\"{u}r Physik, Universit\"{a}t Heidelberg, Im Neuenheimer Feld 227, 69120 Heidelberg, Germany}
\author{Markus K. Oberthaler}
\affiliation{Kirchhoff-Institut f\"{u}r Physik, Universit\"{a}t Heidelberg, Im Neuenheimer Feld 227, 69120 Heidelberg, Germany}

\begin{abstract}
Driven systems are of fundamental scientific interest, as they can exhibit properties that are radically different from the same system at equilibrium.
In certain cases, long-lived states of driven matter can emerge, which exhibit new material properties.
In this work, we probe the excitation spectrum of an emergent patterned state in a driven superfluid, finding that its response is identical to that of a one-dimensional supersolid.
In order to extract physical quantities that parametrize the observed sound modes, we apply an effective hydrodynamic theory of superfluid smectics, which is agnostic to microscopic processes.
We therefore use the conceptual framework of supersolids to characterize an otherwise dynamic and far-from-equilibrium state.
\end{abstract}

\maketitle

\begin{figure}
\centering
\includegraphics{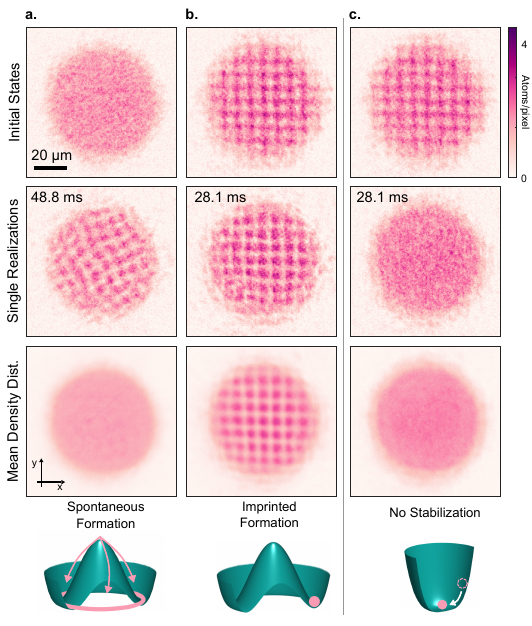}
\caption{\textbf{Initialization of the lattice.} a) Spontaneous formation. The initial state is flat with fluctuations (top), and after a number of drive periods a stable lattice appears, as can be seen in single shots (second row). Averaging the single shots results in a flat density (third row), indicating that the process is spontaneous and reminiscent of dynamics in a Higgs potential.
b) Imprinted Formation. A lattice with a specific orientation and phase is prepared by flashing on a light field, resulting in a lattice density distribution a short time later (first row). At later periods, structures remain remarkably robust (second row), and averaging over many realizations shows the lattice is reproducibly stabilized (third row). This is schematically depicted by preparing the state at a specific point in the Higgs potential. 
c) No stabilization. The lattice preparation procedure can be performed without drive, which shows a quick decay of the contrast of the lattice, and the return to the unmodulated state.}
\label{fig:Fig1}
\end{figure}

An active field of modern research is the periodic driving of system parameters to engineer novel material properties, such as inducing superconductivity \cite{Fava2024} or engineering heat transport \cite{Li2012}.
Though driving generically leads to heating, in certain cases it can produce ordered stationary states, enabling the successful application of mathematical descriptions developed for equilibrium scenarios \cite{Wigner1995,Goldman2014,Oka2019,Kongkhambut2022,Dogra2023}.
Bosonic quantum gases, for instance, have been shown to spontaneously develop self-stabilized, periodic density modulations when the two-particle interaction strength is driven in time \cite{Engels2007_FaradayinBEC, Smits2018_SpaceTimeCrystal,Jason2019, Zhang2020, Dupont2023}.
These patterned states share key physical properties to a seemingly different equilibrium physical system, namely supersolids \cite{Gross1957, Andreev1969, Leonard2017,Li2017,Recati2023,Poli2023,Xiang2024}.

Supersolids are states in which two symmetries, U(1) gauge symmetry and translational symmetry, are separately broken. 
These systems are phase coherent (connected to U(1) gauge symmetry breaking) but also show spontaneously emerging periodic density modulations (translational symmetry breaking), and thus demonstrate an interesting interplay of delocalized particles in localized density structures.
A minimal ansatz for a corresponding order parameter is
\begin{equation}
    \psi = \psi_0\mathrm{e}^{i\theta_{s}}\big[1+\phi\cos\left(k_c x + \theta_{l}\right)\big],
    \label{eq:supersolidmodel}
\end{equation}
where $\psi_0$ is a real constant, $\theta_s$ is the phase of the superfluid, $\phi$ sets the contrast of a periodic density modulation with wavenumber $k_c$, and $\theta_l$ is the spatial phase of the modulation.
Spontaneous breaking of symmetries implies that the groundstates of Hamiltonians that describe supersolids are characterized by finite values of $\psi_0$, $\phi$, and $k_c$, but allow for phases $\theta_{s/l}\in (-\pi,\pi]$.
Global transformations of the phases come at no energy cost. 
Spatial variations, $\theta_{s/l} \rightarrow \theta_{s/l}(x)$, are associated with energy costs $\rho_s(\nabla\theta_{s}(x))^2$ and $B(\nabla\theta_{l}(x))^2$, where the superfluid density $\rho_s$ and the effective layer compression modulus $B$ are the generalized rigidities associated with the two broken symmetries.
These energy costs lead to dynamics that can be understood as coherent excitations of the corresponding Goldstone modes.
Therefore, superfluid phonons (deformations of $\theta_s$) and lattice phonons (deformations of $\theta_l$) can have distinct propagation speeds.

A number of platforms have experimentally demonstrated states that are superfluid and show spontaneously emerging crystalline order.
Originally observed in Bose-Einstein condensates (BEC) inside optical cavities \cite{Leonard2017} and spin-orbit coupled BECs \cite{Li2017}, supersolids have been most extensively probed in gases of atoms with large magnetic dipole moments \cite{Tanzi2019,Chomaz2019,Bottcher2019,Norcia2021}, where a wealth of features related to the superfluidity of the system have been demonstrated \cite{Guo2019,Ilzhofer2021,Tanzi2021,Norcia2022,Biagioni2024,Casotti2024}.

A distinct feature of supersolid systems is that two sound modes arise, which are determined by a finite compressibility of the lattice as well as a modified superfluid fraction \cite{Hofmann_2021,Blakie2023, Sindik2024}.
In dipolar supersolids, experiments have been performed that demonstrate dual frequency response \cite{Natale2019,Tanzi2019Symmetry}, as well as a reduced superfluid fraction \cite{Biagioni2024}.
Recently, two distinct excitation modes have been observed in spin-orbit coupled supersolids \cite{chisholm2024}.
Theoretical studies have developed a hydrodynamic framework constructed using fundamental assumptions about the conserved quantities and (broken) symmetries to describe the excitation branches of supersolids \cite{Yoo2010,Hofmann_2021}.
However, the experimental observation of sound propagation and the extraction of the dispersion relation have proven challenging, due to limited system sizes as well as the use of harmonic trapping configurations.

Recently, we have experimentally demonstrated that driven superfluids can support two-dimensional, self-stabilized periodic density modulations \cite{liebster2024}.
In this system, density modulations emerge due to the growth of occupations at a specific momentum scale set by the drive frequency (\cref{fig:Fig1}a) \cite{Engels2007_FaradayinBEC,Zhang2020,Dupont2023}, and can lead to Faraday patterns in the regime of large occupations and strong nonlinearities \cite{Staliunas2002,Smits2018_SpaceTimeCrystal,liebster2024}.
Theoretically, the order parameter of the state can be described as a stable fixed point solution of the driven Gross-Pitaevskii equation (GPE), indicating that patterns at a finite contrast are long-lived despite the continued drive \cite{Kagan2007,Keisuke}.
This meta-stable nature motivates a characterization of the emergent state via its excitation spectrum.

In this work, we demonstrate that such a meta-stable patterned state of a driven superfluid supports the sound modes of a one-dimensional supersolid. 
By probing the linear response of the system in the hydrodynamic limit, we find universal features of superfluid smectics, i.e. states with spontaneously broken gauge and translational symmetry along a single direction \cite{Yoo2010,Hofmann_2021,Sindik2024}.
We also experimentally probe fundamentally two-dimensional transverse excitations, and find that there is no propagating sound but rather diffusive behavior, indicating that the square lattice structures observed can be described as two independent superfluid smectics in the current parameter regime.

\begin{figure*}
\centering
\includegraphics{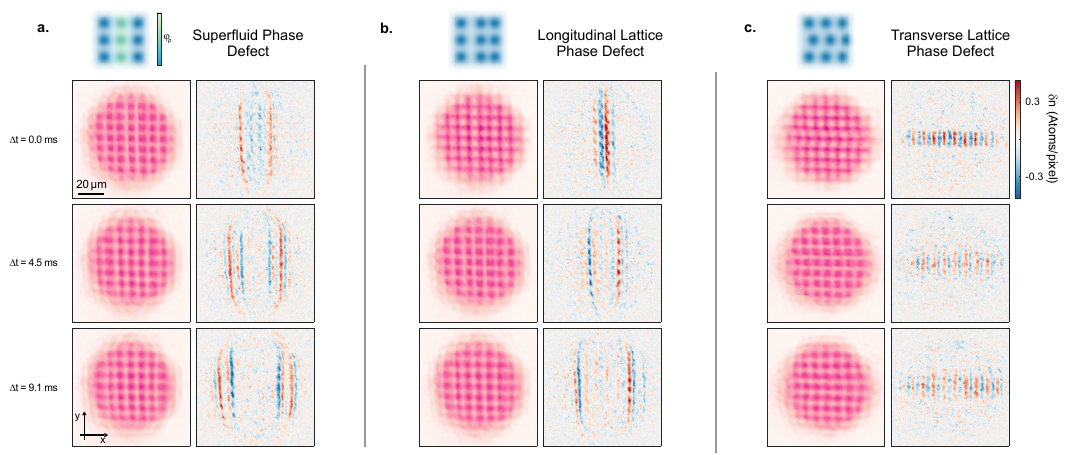}
\caption{\textbf{Wavepacket Propagation.} a) Deformation of superfluid phase. Superfluid flow is induced by locally shifting the phase of the superfluid, by briefly applying a local potential. Density distributions (left column) and density differences relative to unperturbed lattices (right) are shown, indicating the propagation of the wavepacket. In lab time, $\Delta t = 0$ corresponds to 5.1\,ms. b) Longitudinal lattice deformation. By preparing lattices with imperfections, we can probe their propagation. If the central column is shifted horizontally, it initiates two counter-propagating lattice compression wavepackets. Density (left) and density difference plots (right) show the propagation outwards, and the clear restoration of the lattice after the wavepacket has passed. Remaining structure in the density difference is a result of reduced contrast in the perturbed case. In lab time, $\Delta t = 0$ corresponds to 0.6\,ms. c) Transverse deformation. The wavepacket is prepared similarly, but here with a $y$-dependent shift of the standing wave in $x$. The density and density difference plots show a slow diffusion of the initial excitation. In lab time, $\Delta t = 0$ corresponds to 0.6\,ms.}
\label{fig:Fig2}
\end{figure*}

\textbf{Experimental Platform} 
In the experiment, we load a BEC of $\sim\,$35,000 potassium atoms into a trap with a two-dimensional geometry, where the z-axis is tightly confined with a repulsive optical lattice (532\,nm) with trap frequency $\omega_z = 2\pi\times 1.5\,\mathrm{kHz}$.
In the radial direction, the light-shift potential is shaped with a digital micromirror device (DMD), which we use to create a potential that is flat in the central region and is linearly sloped at the edges (see Methods). 
We drive the scattering length (interaction strength) using the broad \mbox{Feshbach} resonance at 561\,G \cite{HadziFeshbach}, with the form $a_s(t) = \bar{a}\left(1-r \sin(\omega_d t)\right)$, with $\bar{a} = 200\,a_\mathrm{B}$ where $a_\mathrm{B}$ is the Bohr radius, $r = 0.3$, and $\omega_d = 2\pi\times440$\,Hz.
The chemical potential without drive and density modulations is $\mu_0 =  2\pi\times300$\,Hz. 
We take \textit{in-situ} images of the density distribution at high fields \cite{Hans2021}.

For spontaneously emerging patterns, we abruptly switch on the drive (i.e. modulation of the scattering length), and, after 20-30 drive periods, each realization yields a pattern with reproducible contrast and wavenumber, but with a random orientation and lattice phase; the spontaneity of the process is confirmed by the smooth density distribution after averaging many realizations (\cref{fig:Fig1}a).

In order to experimentally probe the excitations of the patterned state, we imprint a specific lattice phase and orientation, which initializes patterns at the beginning of the drive, as shown in \cref{fig:Fig1}b (see Methods for details on state preparation).
Even after 12 periods (28.1\,ms), the lattice remains at a finite contrast and the system shows minimal heating, enabling us to recover the lattice modulation in the mean density distribution.
Preparation of the initial state is optimized experimentally to minimize changes in contrast and wavenumber throughout dynamics (see Methods).
Lattices are imprinted with a wavenumber $k_c = 2\pi/8\,\text{µm}^{-1}$.
In the driven system, the amplitude $\phi$ in the ansatz \cref{eq:supersolidmodel} is complex, and rotates in the complex plane with half the drive frequency (Methods).
This rotation means that stable, maximal density modulations occur after a quarter of each drive period, and half a period later the contrast is minimal. 
We therefore image the cloud stroboscopically at $n_P+0.25$ drive periods, where $n_P$ is an integer, for all quantitative measurements shown in this work.
See \cref{fig:Supp_StatePrep} for a finer time resolution of a single drive period.

When preparing the lattice, we take advantage of the rotation of $\phi$ by imprinting the pattern in the superfluid phase, rather than the density. 
To do so, we quickly flash on a periodically modulated light-shift potential, which locally changes the time evolution of the superfluid phase.
This scheme leads to enhanced lifetimes relative to imprinting the lattice directly in the density of the condensate.
If we imprint a lattice without the stabilizing drive, we see a quick decay of the lattice and the reemergence of the homogeneous density distribution, as shown in \cref{fig:Fig1}c.

\begin{figure*}
\centering
\includegraphics{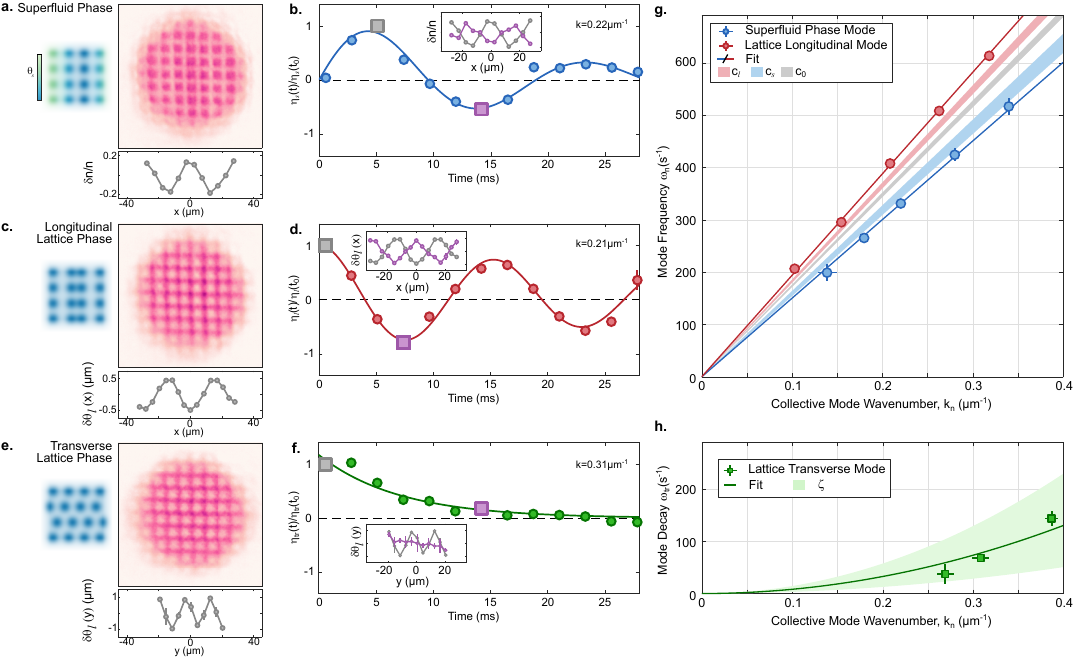}
\caption{\textbf{Hydrodynamic excitations.} a) Superfluid phase modulation. Long-wavelength excitations of the superfluid phase are prepared, indicated by the schematic on the left. A sample mean density distribution two periods after the phase imprint is shown, with the extracted background density deviation $\delta n(x)$ plotted below (gray line). b) The correlator $\eta_s(t)$ of the background density with the reference state, with reference (grey) and inverted (purple) density deviations plotted in the inset. The solid blue line is a fit to the data.
c) Lattice compression modes. A sample density distribution is shown, as well as the extracted lattice deformation $\delta\theta_l (x)$. 
d) Evolution of the correlator for the lattice mode.
e) Transverse lattice perturbations. The phase of the stripe in the $x$-direction is perturbed as a function of $y$. 
f) The correlator shows an exponential decay. The highly structured initial state evolves into a roughly flat distribution with large disorder (purple inset).
g) Extracted dispersion relations. The circles show extracted wavelengths and frequencies of the stable compression modes, whereas the squares show the decay rate of the transverse modes. The shaded regions are the extracted wavepacket speeds from \cref{fig:Fig2}, and the solid lines are linear and quadratic fits to the data.
Standard errors and $1\sigma$ fit errors are either shown or are smaller than the markers.}
\label{fig:Fig3}
\end{figure*}

\textbf{Wavepackets} Identifying distinct, propagating sound modes provides insight into the linear excitation spectrum.
With the capabilities of our experimental platform, we can probe the rigidity of the superfluid and lattice phases to perturbations by writing localized deformations into the phases $\theta_{s/l}(x,y) = \bar{\theta}_{s/l} + \delta\theta_{s/l}(x,y)$, with deformation fields $\delta\theta_{s/l}(x,y)$.

To probe the rigidity of the superfluid phase $\theta_s$, we prepare the crystalline state as described previously, and then pulse on an additional light-shift potential in the central region of the condensate after a half period of driving, resulting in a superfluid phase deformation of approximately the form $\delta\theta_s(|x|\!<\!4\,\text{µm})\!\sim\!0.1\pi$. 
Though we cannot directly access the superfluid phase, the rapid evolution between the region of elevated phase and the remainder of the cloud induces superfluid flow, leading to density perturbations, which can be directly observed.

Density distributions after the phase imprint can be seen in the left column of \cref{fig:Fig2}a, where over and underdensities can be seen perturbing the lattice.
The wavepackets are more clearly seen in density difference plots (right), where the density distributions of perturbed lattices are subtracted from ones prepared in the same way but without a perturbation.
In order to extract a speed of sound, we integrate the density difference signal in a central 9.5\,µm wide region of the condensate.
We track the distance between the two counter-propagating underdensities to determine speeds, which reduces systematic effects like slow sloshing of the background.
The times are selected such that different wavepackets are at comparable positions in the cloud.
The extracted speed of sound is $c_s = 1.59(4)$\,µm/ms, which is reduced from the initial state without modulation and drive of $c_0 = 1.74(2)$\,µm/ms (Methods).

To probe the rigidity of the lattice phase, we imprint lattices with defects, shown in \cref{fig:Fig2}b.
Longitudinal compression sound waves are induced by shifting one column horizontally, with the approximate form $\delta\theta_l(|x|\!<\!4\,\text{µm})\!\sim\!0.4\pi$; for a finite layer compressibility, this lattice phase kink should come at an energy cost, and will induce the propagation of lattice wavepackets through the system.
In the lower half of \cref{fig:Fig2}b, we show both density and density difference plots that clearly show this propagating through the lattice, indicating the presence of longitudinal sound.
After the lattice wavepackets have moved through the system, the phase of the lattice is restored to that set by the bulk $\bar{\theta}_l$, and remaining structure observed in the density difference is due to slightly reduced contrast of the lattice relative to the reference.
By tracking the propagation of the maximum and minimum of the wavepacket, we extract a speed of sound for longitudinal lattice phonons $c_{l} = 1.83(3)$ µm/ms, notably faster than the superfluid phonons, indicative of a separate branch of the dispersion relation.

Finally, a transverse lattice wavepacket is prepared similarly. 
Here, the central row is shifted horizontally as a function of the vertical coordinate, $\delta\theta_l(|y|\!<\!4\,\text{µm})\!\sim\!0.5\pi$ (\cref{fig:Fig2}c). 
In the dynamics, we see that the initial sharp jumps in the deformation field are smoothed, and the initially localized deformation broadens.
A Gaussian fit to the wavepacket shows that its squared width initially increases linearly in time, characteristic of a diffusive mode (Methods).

\begin{figure}
\centering
\includegraphics{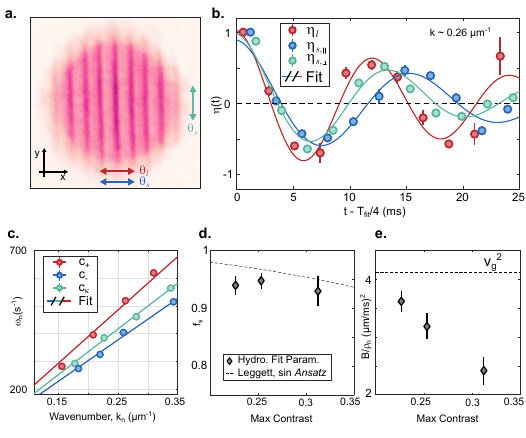}
\caption{\textbf{Determination of parameters.} a) Stripe patterns enable the extraction of layer response, modified superfluid response, and bulk compressibility. 
b) Quantification of response. The points are extracted correlations $\eta(t)$ with the reference density difference. Solid lines are fits. For better comparison of the oscillation, the data and fits are horizontally shifted such that the maximum value of the fit is at $t=0$. c) Extracted dispersion relations, with linear fits. d) Extracted superfluid fraction for three stripe contrasts. e) Extracted layer compressibility modulus for three stripe contrasts. Standard errors and 1$\sigma$ fit errors are either shown or are smaller than the markers.}
\label{fig:Fig4}
\end{figure}

\textbf{Hydrodynamic Perturbations} 
Having identified two propagating and one diffusive mode, we now turn to a more quantitative description of the excitation branches, namely long-wavelength, hydrodynamic perturbations of the system.
These long-wavelength modes can be experimentally challenging to probe in finite sized systems, and here the effectively open boundary conditions implemented through the slanted wall potential enable such experiments (Methods).
Low-energy modes can be excited in the superfluid through long-wavelength phase imprints of the form, $\delta\theta_s(x) \propto \sin(k_nx)$, where $k_n < k_c$.
This induces flow of the background density, as is shown in \cref{fig:Fig3}a, where long-wavelength over- and underdensities are apparent.
Variations of the background density are extracted by subtracting the perturbed lattice from a reference lattice and binning the resulting density difference (bin width of 5\,µm, Methods).
Binning reduces noise in the profiles, but does not result in significant quantitative differences.
To quantify the time evolution, we calculate the time-time correlations of the reference state with other times, defined as $\eta_s(t) = \sum_i\delta n_{i}(t) \delta n_{i}(t_0)$, where $i$ is the index of the bin and runs over the whole cloud, and $\delta n(t_0)$ is the reference state.
Because the perturbations are written into the phase of the superfluid, the initial time shows no density perturbation; the reference is therefore selected as the first time where the density contrast is maximal. 
Figure\,\ref{fig:Fig3}b shows the evolution of $\eta_s$, fitted with a damped sine, which we use to extract the frequency. 
The spatial wavenumber is extracted with a fit to the reference state.

Analogously, compression modes of the lattice are described by slow perturbations of the lattice phase, $\delta\theta_l(x) \propto \sin(k_nx)$.
To track the dynamics, we integrate out the vertical direction and extract the positions of the maxima and minima of the lattice, and plot these relative to an unperturbed lattice, effectively measuring the deformation field $\delta\theta_l(x)$ (\cref{fig:Fig3}c).
The perturbations are small, with displacements on the order of 0.5$\,$µm compared to the lattice spacing of 8$\,$µm, to stay in the regime of linear response.
The time-time correlator is given by $\eta_l(t) = \sum_i \delta\theta_{l}(x_i,t)\delta\theta_{l}(x_i,t_0)$, where $\delta\theta_{l}$ is evaluated at the $i^{\mathrm{th}}$ lattice site, shown in \cref{fig:Fig3}d.
After the imprint, the collective lattice oscillations commence, as can be seen in the inversion of the initially seeded lattice positions (purple curve, inset).

Plotting the wavenumbers and frequencies of these collective modes, we can map out two branches of the dispersion relation, where one is dominated by compressions of the superfluid, and the other by compressions of the lattice, as plotted in  red and blue in \cref{fig:Fig3}g.
We perform linear fits of the form $\omega_n = c_ik_n$, extracting slopes $c_+ = 1.94 (2)\,$µm/ms and $c_- = 1.50(1)\,$µm/ms.

Though \textit{a priori} wavepackets are not expected to move with the phase velocity of long-wavelength excitations, the slope of the measured dispersion agrees remarkably well to the independently measured speeds of lattice and superfluid wavepackets ($c_l$ and $c_s$, shaded regions).
The reduced splitting of the speeds could be explained in that the natural modes underlying the dispersion generically have contributions of contrast variations combined with superfluid and lattice phase modulations \cite{Yoo2010,Hofmann_2021,Sindik2024,Hertkorn2024}.

Finally, we can also probe transverse sound modes, by shifting the lattice horizontally as a function of the vertical coordinate (\cref{fig:Fig3}e), quantified as $\eta_{tr}(t)\,=\,\sum_i \delta\theta_{l}(y_i,t)\delta\theta_{l}(y_i,t_0)$, where $\delta\theta_{l}$ is evaluated at the $i^{\text{th}}$ lattice site.
In contrast to the longitudinal sound of the lattice and superfluid, this excitation does not show clear inversion or oscillation and simply decays into a roughly square lattice with long wavelength distortions, as indicated by the large error bars in the final state (inset \cref{fig:Fig3}f).
No oscillation is apparent, and the data is well described by an exponential decay, with a length-scale dependent decay rate.
The behavior of the collective modes matches with that of the wavepacket, which can now be identified to have a diffusive shape, with a branch of the dispersion of the form $\omega_{tr} = -i\zeta k_n^2$, where $\zeta = 800(130)\,\frac{\text{µm}^2}{\text{s}} = 0.5(1)\,\frac{\hbar}{m}$ is the kinematic viscosity, with $\hbar$ the reduced Planck's constant and $m$ the mass of potassium-39.
Fitting a parabola to the decay rates to extract $\zeta$ gives good agreement to the initial rate of widening of the transverse wavepacket (shaded area).

\textbf{Hydrodynamic Description} The splitting of the dispersion relation into a fast density branch and a slow phase branch is a universal feature of supersolids \cite{Martin1972,Yoo2010,Hofmann_2021,Blakie2023,Sindik2024}.
Long wavelength behavior in thermodynamic phases can be described without a microscopic description of how these modulations arise \cite{Martin1972}.
Though stable states of driven systems are not in general thermodynamic phases, we experimentally observe that energy and atom number are conserved, and the conservation of momentum is upheld because no modulated external potential is imposed. This motivates the application of a hydrodynamic description of the observed dynamics \cite{Yoo2010,Hofmann_2021}.

Along the density modulation of a supersolid, the two gapless, linear branches of the dispersion have slopes $c_\pm$ given by
\begin{equation}
    c_{\pm}^2 = \frac{K}{2\rho} +\frac{B}{2\rho_n} \pm \frac{1}{2}\left[\left(\frac{K}{\rho} +\frac{B}{\rho_n}\right)^2 - 4f_s\frac{KB}{\rho\rho_n}\right]^\frac{1}{2},
    \label{eq:speeds}
\end{equation}
where $\rho = \rho_s + \rho_n$, with $\rho_n$ the normal component density and $\rho_s$ the superfluid density, $f_s$ is the superfluid fraction, $K$ is the bulk compression modulus, and $B$ is the layer compression modulus \cite{Hofmann_2021}.
If the bulk compression modulus of the gas is determined, one can then use the slopes of the measured dispersions to extract the remaining parameters $\frac{B}{\rho_n}$ and $f_s$.

One advantage of our system is that we can stabilize stripes for sufficiently long durations, allowing us to experimentally determine the superfluid response perpendicularly and parallel to the density modulation. 
This gives direct access to the compressibility of the bulk ($\frac{K}{\rho}$) as well as the modified superfluid response ($c_-$) \cite{Hofmann_2021}.
In order to prepare stripe patterns, we repeat the procedure described previously, but now imprint the structure using a light-shift potential that is modulated only along the x-direction.
Though these stripe patterns will develop into square lattices for long times \cite{Keisuke}, short time dynamics do not show significant structure formation in the vertical direction.
As with the lattices described above, we extract the response of the density and phase branches in the $x$-direction, and compare their response to the compressibility in the $y$-axis.
The results, summarized in \cref{fig:Fig4}a-c are consistent with the expected dynamics \cite{Hofmann_2021}.
The excitation branches in the modulated direction split around  the bulk compressibility ($c_\kappa = \sqrt{K/\rho} = 1.66(2)\,\text{µm/ms}$), resulting in a fast and slow branch ($c_+\!=\!1.94(5)\,\text{µm/ms}$ and  $c_-\!=\!1.50(2)\,\text{µm/ms}$).
Using these three slopes, we extract values for the superfluid fraction $f_s=94(1)\%$ and layer compressibility modulus $\frac{B}{\rho_n}=3.3(2)\,(\text{µm/ms})^2$.

For a systematic investigation of the dynamics, we study the excitation velocities for different modulation depths, to probe the emergence of supersolid sound from an unmodulated superfluid. 
By varying the driving amplitudes ($r=0.22, 0.3, 0.4$), different contrasts are stabilized, and dispersions are extracted in each case. 
Contrast is quantified using the integrated one-dimensional density in the central region of the cloud, with contrast values of $\frac{\text{max}(n)-\text{min}(n)}{\text{max}(n)+\text{min}(n)} = 0.23, 0.25, 0.31$.
The extracted superfluid fraction (\cref{fig:Fig4}d) remains constant within error bars, and is compared to the Leggett prediction for a sinusoidal modulation with a given contrast \cite{Leggett1970,Sindik2024}.
The small deviation from the Leggett prediction is remarkable, considering that effects from oscillating phase and density modulations or an enhanced normal component due to directional redistribution of momenta are not considered.
The layer compressibility modulus, however, (\cref{fig:Fig4}e) decreases significantly. 
For comparison, we show $v_g^2$, where $v_g$ is the group velocity at $k_c$, which is the natural response of a superfluid for perturbations in the vicinity of $k_c$.

We note that in contrast to the recent experiments demonstrating anisotropic superfluid fraction in systems with externally imposed lattices \cite{Chauveau2023,Tao2023}, the layer compressibility significantly contributes to the lower branch of the dispersion, i.e. $c_-\neq\sqrt{f_s}c_\kappa$, and therefore shares key features with typical supersolid systems.

\textbf{Outlook}
In this work we have presented the propagation of supersolid sound in a driven superfluid, made possible due to the unique control over the experimental apparatus.
This platform enables the study of a wealth of questions relevant for patterned superfluids, due to the ability to generate systems with many lattice sites, the straightforward preparation of topological defects as well as an extension to three dimensions.
Additionally, a budding field of research is the study of phase transitions out of equilibrium \cite{Fruchart2021}, and our platform provides an ideal testbed for the investigation of such dynamics.
For example, observing the formation of density patterns near the critical temperature \cite{Baena2023}, or the analysis of crystal domains could prove interesting for studying crystallization in driven systems.
Finally, large occupations of excitations that modify properties of the background superfluid open new perspectives of metric engineering, going beyond standard cosmological models \cite{Viermann2022}.

\textbf{Acknowledgements} The authors thank Wilhelm Zwerger for insightful comments on the hydrodynamic description. They also thank Sandro Stringari, Alessio Recati, Blair Blakie, Augusto Smerzi, Luca Pezz\`e, and Ralf Klemt for positive and enlightening feedback, as well as Nicol\`o Antolini, Niklas Rasch, Keisuke Fujii, and Tilman Enss for discussions. This work is supported by the Deutsche Forschungsgemeinschaft (DFG, German Research Foundation) under Germany’s Excellence Strategy EXC 2181/1 - 390900948 (the Heidelberg STRUCTURES Excellence Cluster), under SFB 1225 ISOQUANT - 273811115, and within the QuantERA II Programme that has received funding from the European Union’s Horizon 2020 research and innovation programme under Grant Agreement No 101017733 as well as the DFG under project number 499183856. N.L. acknowledges support by the Studienstiftung des Deutschen Volkes.

\newpage
\bibliography{references.bib}

\section*{Methods}
\setcounter{figure}{0}
\renewcommand{\thefigure}{S\arabic{figure}}

\textbf{Trapping potential} We use an unconventional trapping potential, which is homogeneous in the middle and linearly slanted at the edges:
\begin{equation}
V(\mathbf{x}) =\begin{cases}
          0 \quad & \, |\mathbf{x}| < R_{\mathrm{trap}} \\
        \beta (|\mathbf{x}| - R_{\mathrm{trap}}) \quad & \,|\mathbf{x}|\geq R_{\mathrm{trap}} \\
     \end{cases}
\end{equation}
with $\beta = 2\pi\times18\text{\,Hz/µm}$ and $R_{\mathrm{trap}} = 31$µm.
This specific trap shape suppresses coherent reflections of phonons at the boundary while conserving atom number.
A combination of slowing of the wavefront due to the gradual decrease of the density at the edge, as well as roughness of the potential contribute to the scrambling of reflected wavefronts.

The experiments presented in the main text have also been performed in square traps with both slanted walls and very 'steep' trapping barriers, i.e. $\beta\gg0$.
In square traps with steep barriers, lifetimes are significantly reduced, as even slight roughness or imperfect matching of the wavelength of the excitation to the box size results in reflections that diminish the contrast of the pattern.
Square traps with soft walls result in qualitatively similar results as those described in the main text.

\begin{figure*}
\centering
\includegraphics{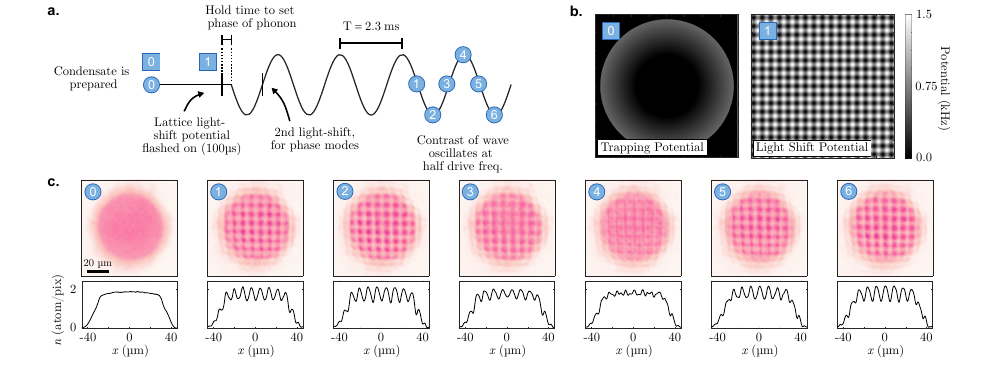}
\caption{a) The experimental sequence. The oscillating line represents the scattering length. We start with a homogeneous system, and then flash on a light shift potential to imprint the lattice. A brief hold time sets the correct phase of the lattice amplitude $R$ relative to the driving. After half a period, if superfluid phase modes are investigated, the second light shift is flashed on. Initial lattice phase deformations for excitation of the corresponding modes are included in the lattice imprint. b) The potentials used for the trap (left) and the light shift for the lattice (right). c) Mean densities throughout the drive. The contrast of the lattice oscillates, as described in \cref{eq:Density}. The curves are horizontal cuts, averaged over a region of 16\,µm in the center of the cloud.}
\label{fig:Supp_StatePrep}
\end{figure*}

\begin{figure}
\centering
\includegraphics{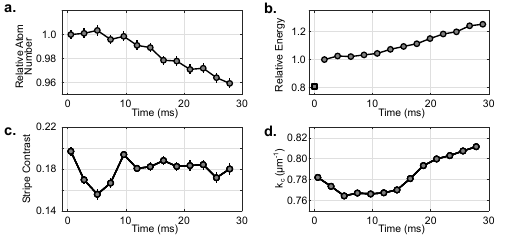}
\caption{a) Atom number varies less than 5\% throughout dynamics. 
b) Energy in system as a function of time, extracted using the measured momentum space distribution $n(k)$, and integrating $\int dk\, n(k)k^2$. The square at $t=0$ is the value without an imprint and driving, and circles are after the imprint, measured stroboscopically at the point when the kinetic energy is maximal. 
c) Contrast of the stripe in $x$ in the central region of the cloud, $C = n(x)_{\text{max}} - n(x)_{\text{min}} / n(x)_{\text{max}} + n(x)_{\text{min}}$, extracted using a sine-fit to mean density distributions of the unperturbed system.
d) Wavenumber of the lattice in $x$, extracted using a Fourier transform of mean density distributions. One-sigma fit errors and standard errors are either shown or are smaller than the markers.
\label{fig:Supp_ConservedQuantities}
}
\end{figure}

\textbf{State preparation} After imprinting the lattice patterns and switching on the drive, the system naturally adapts the stripe wavelength, phase of the phonon amplitude $R$, and contrast to the stable solutions of the driven system. 
We tune drive frequency, amplitude, and phase lag, and imprint depth to minimize changes in contrast and wavenumber of the lattice throughout driving, as well as to maximize its lifetime. 
In particular we experimentally determine a hold time of $0.15$ periods before switching on the drive to yield the correct phase of the oscillation of $\phi$ relative to the drive.
The atom number, integrated energy, stripe contrast, and wavenumber of the pattern are shown in \cref{fig:Supp_ConservedQuantities}, and show minimal changes over the relevant timescale of the experiment.

We estimate the imprinted superfluid phase deformations for the seeded state preparation as well as the wave packet initialization as follows.
We use the linear slope of the equilibrium density distribution at the potential edge to determine the maximal light-shift in the plane of the atoms at the typical input intensity, $V_\text{max} = 2\pi\times3.8$\,kHz, achieved if all mirrors of the DMD are switched on.
For the state preparation, we imprint the lattice with a maximal potential of $40\,\%V_\text{max}$ over $100\,\mu$s, corresponding to a phase of $0.3\pi$.
For the wavepacket propagation we imprint $12.5\%V_\text{max}$ over $100\,\mu$s, corresponding to a phase of $0.1\pi$.

\begin{figure}
\centering
\includegraphics{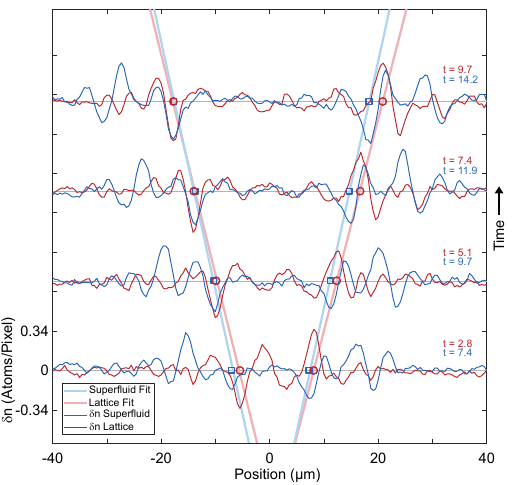}
\caption{Extraction of speeds of sound for superfluid and phase wavepackets. Integrated density differences for superfluid and lattice phase defects are shown for the four times used to fit the speeds. The markers show the extracted position, enabling a comparison to the fit. Lab time of each density difference curve are shown in milliseconds.
\label{fig:Supp_SpeedOfSound}
}
\end{figure}

\textbf{Extraction of Wavepacket Dynamics} To extract wavepacket propagation speeds, we analyze the density contrast in a central region of the cloud,
\begin{equation}
     \delta n(x) = \langle \delta n_{{\text{pert}}}(\mathbf{r}) - \delta n_{{\text{ref}}}(\mathbf{r})\rangle_y,
\end{equation}
where $|y| < 4.5$µm, shown in \cref{fig:Supp_SpeedOfSound}.
In the case of the lattice phonon, we extract the peak position of the wavefronts in $\pm x$ by fitting a parabola to the maximum ($+x$) and minimum ($-x$) of this vector.
For the superfluid phonon, we fit parabolas to the two underdensities.
Due to slight variations of the background density, wavepacket speeds must be compared when the wavepackets are physically in the same region of the cloud.
Because of the differences in preparation, we therefore extract wavepacket positions at different global times, namely between times $t_i \in [2.8\text{ms},9.7\text{ms}]$ for the lattice phonon and slightly later for the superfluid phonon, $t_i \in [7.4\text{ms},14.2\text{ms}]$.
The speed of sound is determined with a linear fit to the distance between the two wavepackets as a function of time.

For the diffusive mode, we perform a similar procedure, but here calculate
\begin{equation}
     |\delta n(y)| = \langle| \delta n_{{\text{pert}}}(\mathbf{r}) - \delta n_{{\text{ref}}}(\mathbf{r})|\rangle_x,
\end{equation}
where $|x| < 4\,$µm, which covers one wavelength of the lattice, using the absolute value to capture positive and negative deviations of the lattice. We perform a Gaussian fit to the quantity $|\delta n(y)|$ of the form
\begin{equation}
     F_g = A \mathrm{e}^{-(y-C)^2/B^2} + D.
\end{equation}
Assuming a Gaussian wavepacket in momentum space and a dispersion of $\omega = -i\zeta k^2$, the squared width of the wavepacket in real space is related to $\zeta$ by
\begin{equation}
     B^2 = 4\left(\sigma_0^2 + \zeta t\right),
\end{equation}
where $\sigma_0$ is the initial width. We fit the times during which this linear growth is apparent, corresponding to the first five periods.

\textbf{Extraction of Collective Mode Oscillations} To extract the deformation field of the lattice phase, we first compute the density contrast of a state with a lattice to a homogeneous density distribution 
\begin{equation}
    \delta n(\mathbf{r}) = \frac{n(\mathbf{r})}{n_0(\mathbf{r})} - 1,
\end{equation}
where $n(\mathbf{r})$ is the patterned state and $n_0(\mathbf{r})$ is the homogeneous state.
We then find the mean density in the vertical direction as a function of $x$ in a region $|\textbf{r}|\!<\!R_{\mathrm{mask}}$,
\begin{equation}
    \delta n(x) = \langle \delta n(\mathbf{r}) \rangle_{y}.
\end{equation}
We fit Gaussians to each maximum and minimum of both the reference and the perturbed lattice, with $R_{\mathrm{mask}} = 36\,$µm such that extracted maxima and minima are in a region of 30$\,$µm from the center.
Lattice displacements $\delta\theta_{l}$ are simply the difference between the position of the perturbed lattice positions to the reference lattice.
Error bars of $\delta\theta_{l}$ are determined using 1-$\sigma$ fit uncertainties of the extracted positions.
For all correlators $\eta_j(t)$, error bars are calculated via Gaussian error propagation.
Fits to the correlator $\eta_l(t)$ are done with the form
\begin{equation}
    F_l(t) = A\mathrm{e}^{-t/B}\cos(2\pi \,C\, t) + D,
\end{equation}
and error bars are given by the $1\sigma$ fit errors to the frequency $C$.
For all fits to determine mode frequencies and decays, the inverse errors of the correlator values are used as weights in the fit.

For superfluid phase deformations, the difference vector $\delta n$ is given by
\begin{equation}
     \delta n(x) = \langle \delta n_{{\text{pert}}}(\mathbf{r}) - \delta n_{{\text{ref}}}(\mathbf{r})\rangle_y,
\end{equation}
with $\mathbf{r} < 30$µm, and is binned in intervals of 5µm.
Error bars of $\delta n(x)$ are determined using a jackknife algorithm to capture shot-to-shot fluctuations.
Fits to the correlator $\eta_s(t)$ are performed using the fit function
\begin{equation}
    F_s(t) = A\mathrm{e}^{-t/B}\sin(2\pi \,C\, t) + D.
\end{equation}
and error bars are given by the $1\sigma$ fit errors to the frequency $C$.

\begin{figure*}
\centering
\includegraphics{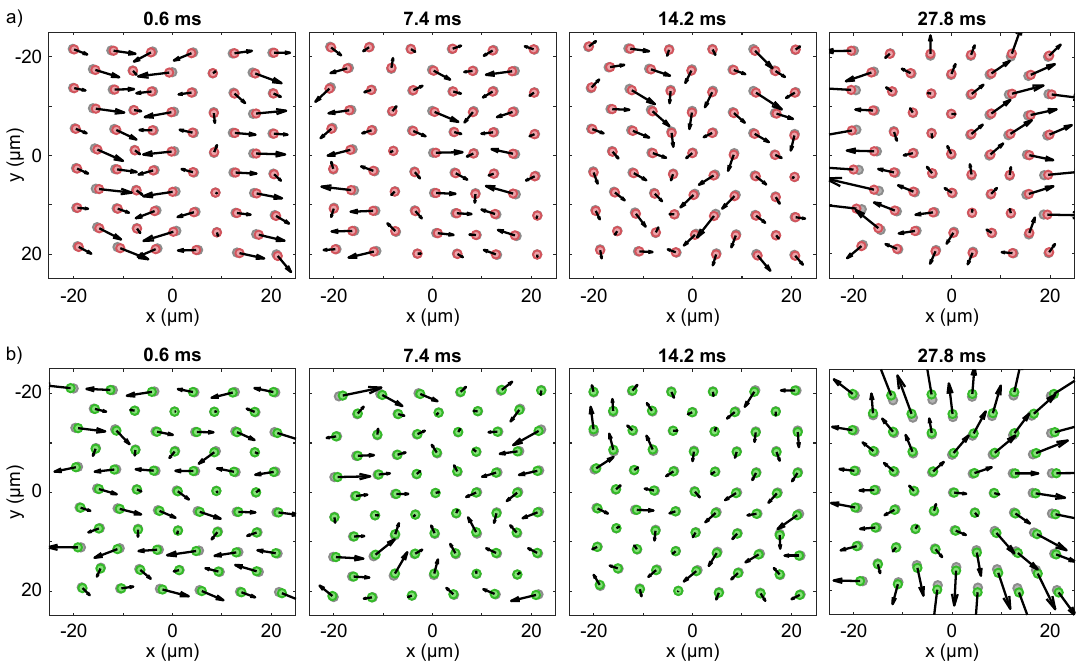}
\caption{\textbf{Two-dimensional lattice deformation fields.} a) Longitudinal lattice mode. The positions of lattice maxima and minima are extracted from a reference (grey) and a perturbed lattice (red). The arrows indicate direction and magnitude of the displacement and are scaled by a factor of 10. For each time, the one-dimensional correlation with the initial state is computed, and depicted in \cref{fig:Fig3}.
b) Collective oscillation extracted in the same way for a transverse lattice mode.\\
}
\label{fig:Supp_QuiverPlots}
\end{figure*}

For transverse modes, we extract the positions of all maxima and minima in two dimensions, as shown in \cref{fig:Supp_QuiverPlots}.
For each row, the displacement relative to the reference is averaged, yielding the deformation field
\begin{equation}
     \delta \theta_l(y_j) = \langle \theta_l{_{\text{pert}}}(x_i,y_j) -  \theta_l{_{\text{ref}}}(x_i,y_j)\rangle_x.
\end{equation}
Error bars of $\delta \theta_l(y_j)$ are calculated using the standard error of the mean horizontal lattice displacement in each row.
A fit of the form
\begin{equation}
    F_{tr}(t) = A\text{e}^{-t/B} + C,
\end{equation}
is performed on $\eta_{tr}$, and decay rates are given by the $1\sigma$ fit errors on parameter $B$.

\textbf{Stabilization mechanism} The stabilization of a specific contrast in the driven system is described in \cite{Keisuke,liebster2024}.
This reduced description captures dynamics of the emergence of the general structure of the pattern but accounts for neither variations in the lattice phase nor modified superfluid response due to the presence of the stripe.
In short, we begin with the driven GPE $i \hbar \frac{\partial \Psi(\mathbf{x},t)}{\partial t} = [ -\frac{\hbar^2 \nabla ^2}{2 m} + V(\mathbf{x}) + g_0\left(1 - r \sin \omega_d t \right) |\Psi(\mathbf{x},t)|^2 ] \Psi(\mathbf{x},t),$ where $m$ is the atomic mass, $\hbar$ is the reduced Planck's constant, and the interaction strength given by $g_0 = \frac{\sqrt{8 \pi} \hbar^2}{m} \frac{\bar{a}_s}{l_z}$, where  $l_z = \sqrt{\frac{\hbar}{m \omega_z}}$.
A minimal model for square lattice patterns is a superposition of two stripes
\begin{equation}
    \label{eq:ansatz}
     \Psi(\mathbf{x},t) = \Psi_{\text{uni}}(t) \Big[1 + \phi_k(t) \cos{(\mathbf{k}\cdot\mathbf{x})} + \phi_p(t) \cos{(\mathbf{p}\cdot\mathbf{x})} \Big],
\end{equation}
with $\Psi_{\text{uni}}(t)$ a uniform, infinitely extended background field with time evolution $\Psi_{\text{uni}}(t) = \sqrt{n_0}\, \exp[-i\mu t - i (\mu / \omega_{d}) r \cos{\omega t}]$, and $n_0$ the 2D density. 
The vectors $\textbf{k}$ and $\textbf{p}$ with $|\textbf{p}| = |\textbf{k}| = k_c$ have an angle $\theta \in [0^\mathrm{o},180^\mathrm{o}]$ between them.
The prefactors $\phi_{k/p}$ are parameterized by Bogoliubov coefficients, and therefore have the form
\begin{equation}
    \label{eq:Phonon}
    \begin{aligned}
    \phi_{k / p}(t) =&\left(1-\frac{\epsilon+2 \mu}{E}\right) R_{k / p}(t) e^{i \frac{\omega_d}{2} t}\\
    &+\left(1+\frac{\epsilon+2 \mu}{E}\right) R_{k / p}^*(t) e^{-i \frac{\omega_d}{2} t},
    \end{aligned}
\end{equation}
where $E = \sqrt{\epsilon ( \epsilon + 2 \mu)}$ and $\epsilon = \frac{\hbar k_c^2}{2 m}$ is the corresponding Bogoliubov energy in units of frequency.
$R_{k/p}$ are amplitudes that vary slowly in time and can be shown to be described in terms of the amplitude equations 
\begin{equation}
    \label{eq:GinzburgLandau}
    \begin{aligned}
    i \frac{d}{d t} R_k(t)=&-i \alpha R_k^*(t) - i\Gamma R_k(t) \\
    &+\lambda\left|R_k(t)\right|^2 R_k(t) \\
    & +\lambda\Big[c_1(\theta)\left|R_p(t)\right|^2 R_k(t)+c_2(\theta) R_p(t)^2 R_k^*(t)\Big].
    \end{aligned}
\end{equation}
Here, $\alpha = r \frac{\mu\epsilon}{2 E}$ and $\Gamma$ is a phenomenological damping constant. 
Other constants $\lambda = \mu \frac{5\epsilon + 3\mu}{E}$, $c_1(\theta)$ and $c_2(\theta)$ are set by $k_c$ and the angle between $\mathbf{k}$ and $\mathbf{p}$. 
Stable solutions of the driven system are given by the fixed-points of these amplitude equations.
At such stable points, the stripe amplitude $\phi$ rotates in the complex plane.
Here, the temporary stabilization of stripes is apparent: if one sets $R_p \sim 0$, a single stripe is self-stabilized, and nonlinear coupling between stripes is negligible.

The oscillation of the lattice contrast (seen in \cref{fig:Supp_StatePrep}) is captured by the density distribution resulting from this \textit{ansatz}, which in one dimension is given by
\begin{equation}
\begin{aligned}
    n(x,t) = &\bar{n}\Bigg(1 + 4 |R|\cos{\left[\frac{\omega_d}{2}t + \varphi\right]}\cos{k x} + \\
    &4 |R|^2\left(1 + 2\frac{\mu}{\epsilon}\sin^2{\left[\frac{\omega_d}{2}t + \varphi\right]}\right)\cos^2k x\Bigg),
    \end{aligned}
    \label{eq:Density}
\end{equation}
where $\varphi$ is the phase of the amplitude $R$, i.e. $R=|R|\exp\{i\varphi\}$.

\textbf{Data Availability} The datasets generated and analyzed for the above study are available from the corresponding author upon reasonable request.

\textbf{Code Availability} The conclusions of this study do not depend on code or algorithms beyond standard numerical evaluations.

\end{document}